# Conductance modulation in graphene nanoribbon under transverse asymmetric electric potential.

S. Bala Kumar,[a)] T. Fujita, and Gengchiau Liang[b)]

Department of Electrical and Computer Engineering, National University of Singapore, Singapore, 117576,

[*]Corresponding author. E-mail: a) brajahari@gmail.com and b) elelg@nus.edu.sg

## Abstract

We theoretically study the effect of transverse electric potentials on the transport properties of armchair graphene nanoribbons (AGNRs), formed by pairs of asymmetric gates placed along the side of the ribbon. Single pair and dual pair configurations are considered. We also examine the effect of hollows (spatial regions void of carbon atoms) in the AGNR channels. We find that the use of hollowed AGNRs in the dual pair configuration allows for a significant modulation of the transport gap, when the two pairs have opposite polarity of gate bias. Furthermore, we show that for the dual-gate system, hollowed AGNR channels exhibit the optimal ratio of ON-state to OFF-state conductance, due to the smaller OFF-state conductance compared with spatially homogenous AGNR channels. Our results indicate that transverse gate technology coupled with careful engineering of hollow geometry may lead to possible applications in graphene-based electronic devices.

## 1. Introduction

The graphene nanoribbon (GNR), a quasi-one dimensional system, has attracted considerable studies due to its unique material and transport properties under electronic[1-5], magnetic[4-13], and optical[11-12] fields. In particular, armchair GNRs (AGNRs) exhibit semiconducting behavior coupled with an extremely low carrier effective mass, making them a potential candidate for novel channel materials in the next generation of field-effect-transistors[14-22]. Furthermore, by varying their shape to create heterostructures, GNRs might be implemented into the applications of quantum dots[23] and resonant-tunneling-diodes[24]. In addition to these traditional device designs, recent studies also investigate the unique properties of GNRs and their possible device applications, such as in magnetoresistive [13] and spintronic devices[25,26]. Recently, it has been theoretically shown that the application of a transverse electric field (E-field) causes a spectral shift of the conduction and valence bands of the AGNR, resulting in an upward (downward) shift for the conductance (valence) band[1-4]. Therefore, the transport gap of an AGNR decreases, especially at a high E-field, and the electron transport properties can be modulated by the transverse E-field. However, this variation is limited and its sensitivity to the gate voltage is very low for conventional device applications[27].

In this paper, we apply transverse-asymmetric-electric-potentials (TAEP) across GNRs [3] as shown in Fig. 1. Unlike conventional top-gated devices, the conductance modulation in this case is achieved by dynamically modifying the bandgap of the GNR channel through the pair of asymmetric side gates, $+V_g$ and $-V_g$. Owing to its planar structure, the channel layers of multiple such devices can be stacked vertically as shown in Fig. 1(a). This allows for multiple channels to be controlled by a single set of gates, allowing for more compact IC integration, such as 3D chips, com-pared to the conventional top gated-FET structure.

However, for this design having a single pair of TAEP gates (single-TAEP), a large E-field is required to close the gap. As a result, it may not be practical for applications. Therefore, we further investigate the effect of a dual-TAEP gate[26] structure as shown in Fig. 1(c), and also introduce a hollow (a spatial region void of carbon atoms) in the AGNR. Compared to homogenous AGNRs, the transport gap of hollowed AGNRs is more sensitive to the variation of the transverse E-field, and can be tuned by careful engineering of the hollow's geometry. Furthermore, unlike homogenous AGNRs, the transport gap of hollowed AGNRs depend on the electronic configuration of the dual-TAEPs. When the two pairs of TAEPs are applied in a parallel (anti-parallel) configuration (corresponding to $V_g>0$ (<0) in Fig. 1(c)), the transport gap is increased (decreased). Due to the significant modulation of the transport gap, the electronic conductance across hollowed AGNRs in dual-TAEP structures can vary considerably. It indicates the possibility to implement these dual gate designs in 3D chip architectures.

**2. Simulation Model**

To investigate the electronic transport in hollowed AGNRs, we use the real space π-orbital tight binding Hamiltonian[28,29], given by $H = \sum_{n} V_n a_n^+ a_n - \sum_{n,m} t_{n,m} a_n^+ a_{m'}$ , where $t_{n,m}$ is the hopping energy between two bonded atoms $m$ and $n$, and $V_n$ is the electrostatic potential at site $n$ caused by the gate potential. The Laplace potential assumed in this work varies across the transverse direction of the AGNR[4]. The electron transport behavior in the GNR is studied by using the non-equilibrium Green's function (NEGF) formalism[30]. In the NEGF formalism, the electron transmission at energy E across the GNR is $T(E) = Trace[\Gamma_S(E) G^r(E) \Gamma_D G^r(E)]^+$ , where $G^r(E) = [EI - H - \Sigma_S(E) - \Sigma_D(E)]^{-1}$ is the retarded Green's function of the GNR channel, $\Sigma_{S(D)}(E)$ is the self energy of the source (drain) leads with semi-infinite homogenous GNRs, and the coupling between the source (drain) and the channel, $\Gamma_{S(D)}(E) = i[\Sigma_{S(D)} - \Sigma^+_{S(D)}]$. For a given Fermi level ($E_F$), the source-drain conductance under a small source-to-drain bias, $G_D(E_F)$, and the electron flux, $j_{nm}(E_F)$, across two atoms $n$ and $m$ are computed as

$$G_D(E_F) = \frac{q^2}{h} \int_{-\infty}^{\infty} T(E) \frac{-\partial f_0(E - E_F)}{\partial E} dE ,$$

$$j_{nm}(E_F) = (q/h) \left[ t_{mn} G^<_{nm} - t_{nm} G^<_{mn} \right],$$

where $G^<_{nm}(E) = i \langle a_m^+ a_n \rangle$ is the lesser Green's function[31,32]. Referring to Fig. 3, $N_y$ is the total number of dimer rows, while $N_x$ is the total number of unit cells. We represent the width (length) of the AGNR with $N_y$ ($N_x$). The width (length) of the hollow is represented by $N_y$' ($N_x$'). We use the notation $n$-AGNR for an AGNR with $N_y = n$.

### 3. Results and Discussion

Firstly, we study the variation of the electronic band structure (E-K dispersion relations) of AGNRs under the single-TAEP as shown in Fig. 2. Previous studies have shown that AGNRs can be classified into three types, i.e. AGNR*3p+1*, AGNR3p and AGNR3p-1, corresponding to the total number of dimers, $N_y$=3p+1, 3p, and 3p-1 [p=1, 2, 3, …] respectively[33]. In this work, semiconducting AGNR3p+1 are selected since these have the largest transport gap when $V_g$=0. Fig. 2(a) shows the change in band structure with $V_g$ for a 13-AGNR. Due to the spectral shift of the conduction and valence band states at the edges [1-4], the minimum separation between the conduction and valence band, i.e. the transport gap $E_g$, changes with increasing $V_g$. In Fig. 2(b), the variation of the transport gap with increasing $V_g$ for AGNRs with different widths is shown. At $V_g=0$, AGNRs with larger width have smaller transport gap. As the $V_g$ increases the transport gap is initially unchanged until it reaches a critical value. Beyond this value the transport gap starts to decrease almost linearly. Under a sufficiently high $V_g$, the gap closes, i.e. $E_g \rightarrow 0$. This indicates a transition of the AGNRs from a semiconducting to metallic state due to the applied TAEP. Fig. 2(b) also shows that the transport gap of wide AGNRs tends to decrease and approach zero at lower $V_g$ compare to narrower AGNRs. The rate at which the gap decreases with $V_g$, i.e. $\Delta E_g/\Delta V_g$, however, is nearly identical for all AGNRs. For all the cases, the decrease in $E_g$ due to $V_g$ is very small, i.e. $E_g/\Delta(eV_g)<1$, indicating that a large bias is needed to close the transport gap, and that it might not be practical in commercial field-effect-transistor applications[27].

Next, we investigate dual-TAEP gated structures. Unlike the single-TAEP case, in a dual-TAEP gate structure we can apply parallel (P) or anti-parallel (AP) E-field configuration by setting $V_g>0$ or $V_g<0$, respectively. In both cases, the left pair of gates always keep the same sign whilst the sign of the right pair varies (see Fig. 3). As a result, when $V_g>0$ the two pairs of gates work under the same fields (defining the P-configuration) while for $V_g<0$, the two pairs work under opposite fields (defining the AP-configuration). To understand the transport properties under these two different configurations clearly, the energy $E=E_F-E_i=0.3eV$ ($E_i=0$) is selected to keep only one conduction subband involved in the carrier transport. When the TAEP is applied in a P-configuration, c.f., Fig. 3(a), under $V_g=1V$, an E-field is produced across the width of the GNRs, which shifts the forward conduction states to the top-edge. As a result, the electrons are transmitted across the top-edge of the GNRs, and the trans-mission remains high, i.e. T=1. When we apply the dual-TAEP in an AP-configuration (Fig. 3(b)), under $V_g$=-1V, the conduction channel splits into two - on the top (bottom) edge for the left (right) side of the channel. Due to the different channels under an AP-configuration, an extremely low transmission might be expected. However, the electrons can still transits from top-edge to bottom-edge via the centre (width wise) of the channel. Due to spatial scattering caused by the electrons flowing from the top edge to the bottom edge, T decreases slightly to 0.76.

To further decrease the transmission, we introduce a potential barrier in the centre of the AGNR by creating a hollow, as shown in Figs. 3(c) and (d). In the hollowed structure, under the P-configuration, the electron transmission remains high, similar to the homogenous structure. This is because electrons are mainly transmitted across the top-edge and thus are not affected by the hollow. However, in the AP-configuration, the hollow resists the flow of electrons from the top to bottom-edge, and thus the transmission is considerably decreased as shown Fig. 3(d), i.e. T=$10^{-15}$. This indicates that compared to the homogenous AGNR, the electron transmission across the hollowed AGNR channel can be considerably modulated by controlling the gate voltage in dual-TAEP gated structures.

To gain a more detailed understanding of the electronic transport in hollowed structures, firstly we investigate T(E) under the different structural conditions, without any transverse E-field. The transport gap is determined as the range of energy where the transmission is suppressed closed to zero, which we set as T<0.01. In a hollowed structure, two AGNR strips (edge-strips) are formed at the top and bottom of the hollow. We assume that the width of the edge-strips are the same for both the strips and given by $N_y''=(N_y-N_y')/2$. Fig. 4(a) and (b) show the T(E) spectrum due to the variation in $N_y''$, for AGNRs with fixed $N_y'$=3 and $N_y$=17, respectively. The dotted lines refer to the T(E) of the homogenous AGNR. For a homogenous AGNR the transport gap is determined by the total width of the AGNR ($N_y$).[33] However, for the hollowed AGNRs, when the length of the hollow is sufficiently large, the edge-strips can be treated as individual channels, and thus the GNR behaves as a heterostructure consisting of: 1) AGNRs with width $N_y$ at the left/right of the hollow, and 2) AGNRs with width $N_y$" at the top/bottom of the hollow. Furthermore, it is also found that, if the transport gap of $N_y$"-AGNR is larger than the transport gap of $N_y$-AGNR, the transport gap is increased. Otherwise, the transport gap is similar with that of the homogenous GNRs, indicating that the effective transport gap is determined by the channel with the largest transport gap. To confirm this hypothesis, we fix the edge-strip widths to $N_y$"=7, such that the strips have a large transport gap. Referring to the circled region in Fig. 4(c), we found, regardless of the value of $N_y$, that the transport gap remains almost constant.

Next, we study the effect of TAEP on the transport gap. The central panel of Fig. 5(a) shows the change in the transport gap for homogenous [Ny=16, Nx=100,] and hollowed structures [Ny=16, Ny'=2, Nx=100, Nx'=90] with varying $V_g$. The left and the right panels show the transmission of homogenous and hollowed structures under $V_g$=-1 (green line), 0 (red line), and 1 V (blue line), respectively, and indicate how to extract Eg/2 corresponding to the central panel. In principle, the transport gap of a homogenous AGNR shows a symmetric variation for P and AP configuration, whereas the transport gap of the hollowed AGNR increases to a maximum value, and then sharply decreases to zero as $V_g$ decreases. This is because of the different physical mechanism underlying the variation of $E_g$ in the two systems. In the homogenous structure, the transport gap variation is predominated by the spectral shift of the conduction and valence band states at the edges [refer to Fig. 2]. However, for the hol-lowed AGNR, the transport-gap variation involves different mechanisms which follow from two unique characteristics 1) the edge-strips form two parallel transport channels, and 2) the potential at the middle of the hollow is zero. Therefore, in the P-configuration, the average potential in the top (bottom) strip is positive (negative), and thus electronic bands are shifted to lower (higher) energy.[4] As Illustrated in Fig. 5(b), this shift decreases the transport gap. However, when $V_g$ is very large, the decrease in the transport gap of a hollowed AGNR is limited by the transport gap of the homogenous AGNR. This is because, in a hollowed AGNR, the left and right strips around the hollow can be considered as semi-infinite homogenous AGNRs, and thus the transport gap is determined by the competition between the edge-strips and the homogenous AGNRs. On the other hand, in the AP-configuration, the average potential along the top/bottom strip varies from positive/negative (along the left-half of the strip) to negative/positive (along the right-half of the strip) as shown in Fig. 5(b). As a result, a junction barrier is formed within the strip, and this barrier increases the effective difference between the conductance and valence band, resulting in an increase in the transport gap with decreasing $V_g$. However, when the transport gap becomes twice the transport gap of the top/bottom GNR strips, any further decrease in $V_g$ causes the gap to drop to zero suddenly, due to the overlap between $E_c$ and $E_v$ of GNR strips which allows electron tunneling through these two bands.

Finally, we compare the source-drain conductance, $G_D$ across AGNRs with varying transverse $V_g$ for both the hollowed and homogenous cases. In Fig. 6(a), we investigate the effect of the TAEP on the $G_D$-$V_g$ curve of homogenous AGNRs of different widths. Unless otherwise state, we set the Fermi level $E_F$ at the intrinsic level (E=0). As the $V_g$ varies, the transport gap changes, resulting in a change of source-drain conductance at the intrinsic level of E=0, i.e. $G_D$ ($E_F$=0). We denote the maximum and minimum $G_D$ as $G_{MAX}$ and $G_{MIN}$, respectively. As $N_y$ increases, the transport gap under $V_g=0$ decreases, resulting in higher $G_{MIN}$. On the other hand, the $G_{MAX}$ remains the same under different $N_y$. It is because when $G_D = G_{MAX}$, the transport gap $E_g \approx 0$ regardless of $N_y$, with one subband for carrier transport for all three cases. Moreover, the $G_{MAX}$ is achieved at lower $V_g$ for devices with larger width, since for larger $N_y$, the transport gap begins decreasing at lower $V_g$ [refer Fig. 2].

Next, Fig. 6(b) shows the effect of the gate on $G_D$-$V_g$ curve of hollowed AGNRs of the different widths. Unlike the homogenous AGNR, the conductance change in the hollowed AGNR is caused by the formation of the gated junction barrier in both of the edge-strips, and thus the conductance changes significantly as $V_g$ varies from AP ($V_g<0$) to P ($V_g>0$) configuration. Referring to the dashed arrows in Fig. 5(b), the electronic transport for $G_{MIN}$ is mainly determined by the electrons tunneling through the potential barrier across the edge-strips. Therefore, when the length of the edge strip, i.e. $N_x$' is longer, the tunneling probability is smaller. This decreases the minimum $G_D$ to a much lower value, resulting in a higher $G_{MAX}/G_{MIN}$ ratio. When $V_g$ is very small (in the AP configuration), there is a steep increase in $G_D$, i.e. a steeper $G_{MAX}$-$G_{MIN}$ transition. Referring to Fig. 5, when $V_g<<0$, band-to-band tunneling occurs within the edge strips, resulting in high transmission around E=0. Due to the band-to-band tunneling, a steep $G_{MAX}$-$G_{MIN}$ transition is obtained in the AP-configuration. Note that, $N_x$'=0 refers to the homogenous structure, where the $G_D$ curve is symmetrical at $V_g=0$. For large $V_g$, the $G_D$ is limited by the $G_D$ of the homogenous structure. Furthermore, we studied the effect of $V_g$ on $G_D$ when we change the Fermi level of the AGNR in Fig. 6(c). The Fermi level can be varied either by doping or by applying a back gate voltage. In general, when the Fermi level is increased: 1) $G_{MIN}$ is higher - the thermal electrons are transported at energy closer to the first con-duction sub-band edge, and thus the thermal current is larger at the $G_{MIN}$, and 2) the $G_{MAX}$ is achieved at lower $V_g$– only a smaller $V_g$ (smaller threshold voltage) is required to increase the electron energy such that it is higher than the first conduction sub-band edge.

**4. Conclusion**

We study the effect of single and dual-TAEP on the electronic transport gap of homogenous and hollowed AGNRs. Although the transport gap of AGNRs decreases under single-TAEPs, the reduction in the gap with $V_g$ is not sufficient for application in commercial electronic devices. To improve the sensitivity of the transport gap variation to the $V_g$, we introduce a dual-TAEP gated structure. This structure allows two different configurations of the transverse E-fields (P and AP-configuration). Although the case for homogeneous GNRs show similar results to the single TAEP system, interestingly, the transport gaps of hollowed AGNRs depends on the configuration of the dual-TAEP, and shows a significant variation with $V_g$. This is because the transport gap of hollowed AGNRs, unlike that of homogenous AGNRs, is a result of four different mechanisms; band-to-band tunneling, junction barrier, band shifting and bandgap modulation by the transverse field, where the former three can considerably affect the electron transport. Finally, we evaluate the source-to-drain conductance of the dual-TAEP structure for hollowed and homogenous AGNRs by varying $N_y$, $N_x$, and $E_F$. We show that compared to the latter, the former can provide a much smaller OFF-state conductance, resulting in the considerably larger ratio of ON-state to OFF-state conductance. Our study indicates the possible applications of transverse gate technology coupled with the careful engineering of hollow geometry in graphene-based electronic devices.

**Acknowledgement**

The computations were performed on the cluster of Computational Nanoelectronics and Nano-device Laboratory, National University of Singapore. This work is supported by A*STAR SERC 0821010023 and National Research Foundation CRP "Graphene and related materials and devices".

**References**
[1] D. S. Novikov, Phys. Rev. Lett. 99, 056802 (2007).
[2] Hassan Raza and Edwin C. Kan   Phys. Rev. B 77, 245434 (2008) .
[3] C. Ritter, S. S. Makler, and A. Latgé, Phys. Rev. B 77, 195443 (2008).
[4] Kausik Majumdar, Kota V. R. M. Murali, Navakanta Bhat and Yu-Ming Lin, Nano Lett. 10, 2857 (2010).
[5] V. Lukose, R. Shankar, and G. Baskaran, Phys. Rev. Lett. 98, 116802 (2007).
[6] C.P. Chang, C.L. Lu, F.L. Shyu, R.B. Chen, Y.C. Huang, M.F. Lin, Physica E 82, 27 (2005).
[7] Y. Shibayama, H. Sato, T. Enoki, M. Endo, Phys. Rev. Lett. 84, 1744 (2000).
[8] S.C. Chen, T.S. Wang, C.H. Lee and M.F. Lin, Phys. Lett. A 372, 5999 (2008).
[9 ] J.Y. Wu, J.H. Ho, Y.H. Lai, T.S. Li and M.F. Lin, Phys. Lett. A 369, 333 (2007).
[10] K. Harigaya, Chem. Phys. Lett. 340, 123 (2001).
[11] K. Wakabayashi, M. Fujita, H. Ajiki, and M. Sigrist, Phys. Rev. B 59, 8271 (1999).
[12] Y.C. Huang, C.P. Chang, M.F. Lin, Nanotechnology 18, 495401 (2007).
[13] S. Bala Kumar, M. B. A. Jalil, S. G. Tan, Gengchiau Liang, J. Appl. Phys. 108, 033709 (2010).
[14] G. Fiori and G. Iannaccone, IEEE Electron Device Lett. 28, 760 (2007).
[15] M. C. Lemme, T.J. Echtermeyer, M. Baus, H. Kurz, IEEE Elec.Dev. Lett. 2007, 28, 282.
[16] B. Obradovic, R. Kotlyar, F. Heinz, P. Matagne, T. Rakshit, M. D. Giles, M. A. Stettler, and D. E. Nikonov, Appl. Phys. Lett. 88, 142102 (2006).
[17] Kai-Tak Lam, Dawei Seah, S. K. Chin, S. Bala Kumar, G. Samudra, Yee-Chia Yeo, and Gengchiau Liang, IEEE Electronic Device Letter, 31(6) 557, 2010.
[18] G.-C. Liang, N. Neophytos, D. Nikonov, and M. Lundstrom, *IEEE Transactions on Electron Devices,* vol. 54 (4) 677 - 682, April 2007.
[20] Y. Ouyang, Y. Yoon, J. K. Fodor, and J. Guo, Appl. Phys. Lett. 89, 203107(2006).
[21] G.-C. Liang, N. Neophytos, D. Nikonov, and M. Lundstrom, Journal of Applied Physics, 102, 054307, 2007.
[22] Xinran Wang, Yijian Ouyang, Xiaolin Li, Hailiang Wang, Jing Guo, Hongjie Dai. Phys. Rev. Lett.   100, 206803 (2008).
[23] Hao Ren, Qun-xiang Li¤, Qin-wei Shi, Jin-long Yang, CHINESE JOURNAL OF CHEMICAL PHYSICS, 20 489 (2007).
[24] H. Teong, Kaitak Lam, B. K. Sharjeel and Gengchiau Liang 2009 J. Appl. Phys. 105 084317.
[25] Young-Woo Son, Marvin L. Cohen, and Steven G. Louie, Nature 444, 347, (2006).
[26] Jing Guo, D. Gunlycke, and C. T. White, Appl. Phys. Lett. 92, 163109 (2008).
[27] S. Bala kumar, T. Fujita, Gengchiau Liang, Proceeding of 2010 Int. Conf. on Solid State Device Mater. (September 2010).
[28] R. Saito, M. Fujita, G. Dresselhaus, and M. S. Dresselhaus, Appl. Phys. Lett. 60, 2204 (1992).
[29] A. H. C. Neto, F. Guinea, N. M. R. Peres, K. S. Novoselov, and A. K. Geim, Rev. Mod. Phys. 81, 109 (2009).
[30] S. Datta, *Quantum Transport: Atom to Transistor*, 2$^{nd}$ edition, Cambridge University Press, Cambridge, MA 2005.
[31] L. Keldysh, Sov. Phys. JETP 20, 1018 (1965).
[32]R. Lake, G. Klimeck, R. C. Bowen, and D. Jovanovic, J. Appl. Phys. 81, 7845 (1997).
[33] Y. Son, M. L. Cohen and S. G. Louie., Phys. Rev. Lett. 97, 216803 (2006)

Figures

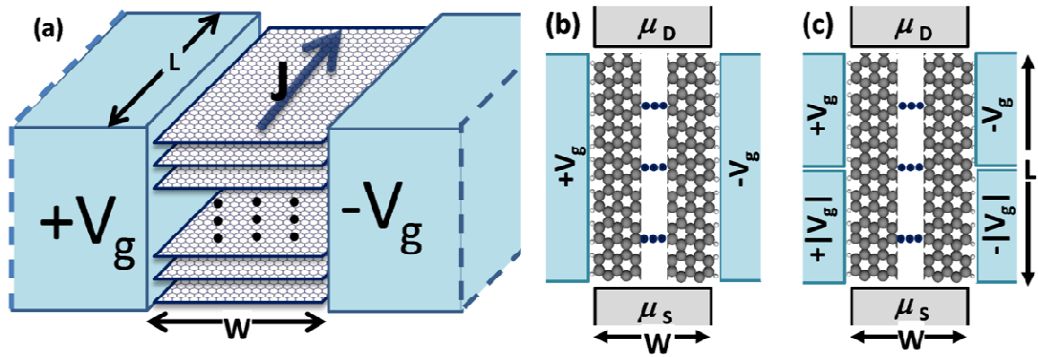

Fig. 1 (a) Structure of a TAEP-gated field-effect-transistor, where multiple AGNR channels are stacked vertically. (b) Top-view of a channel layer - AGNR with a single-TAEP gate. (c) Top view of an AGNR channel layer with dual-TAEP gates. We assume a linear potential drop across the transverse direction of the structure. Semi-infinite AGNR source and drain contacts are used.

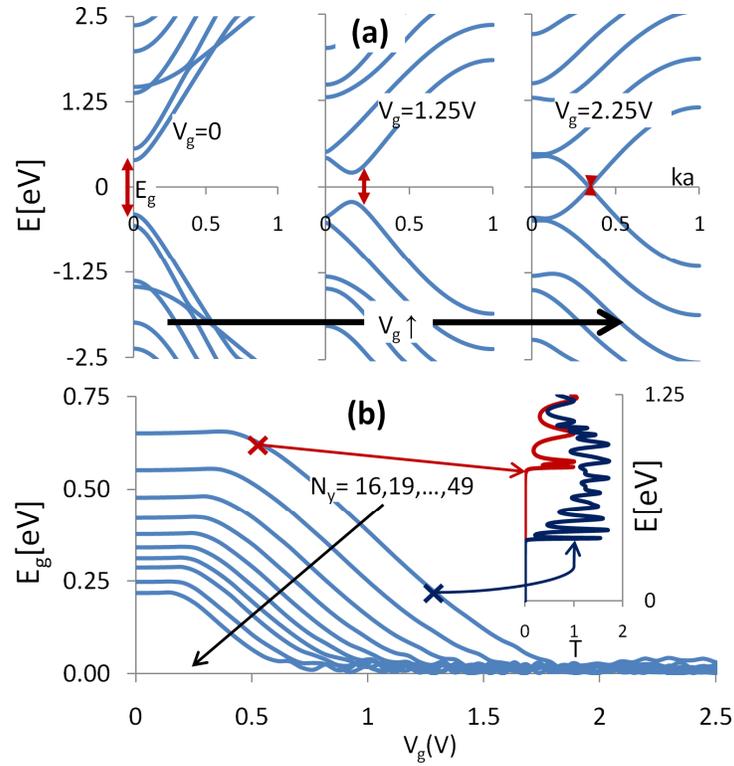

Fig. 2(a) Band structure variation of AGNRs with increasing gate potential, $V_g$. The minimum distance between the valence band and conduction band decreases as $V_g$ increases. [$N_y$=13]. (b) Transport gap variation with increasing $V_g$ for different widths. The inset shows the conductance when $V_g$=0.5V and 1.25V [$N_y$=16].

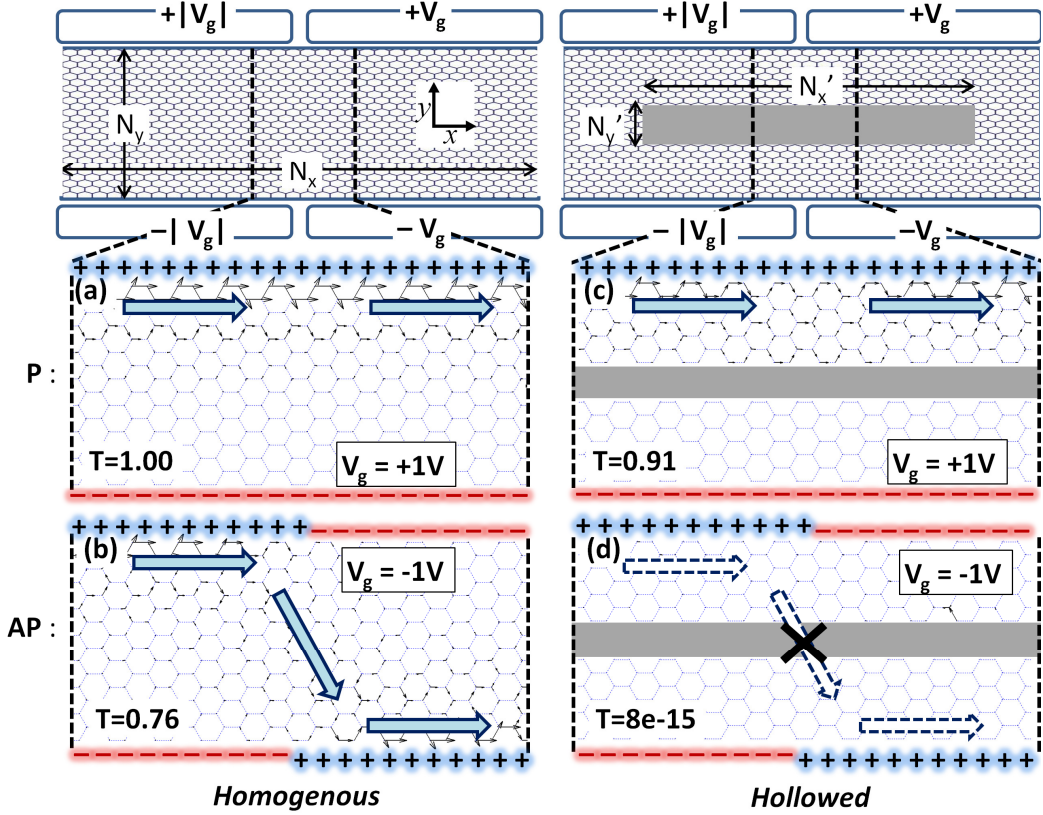

Fig. 3 Electron flux of homogenous and hollowed AGNRs with dual-TAEP gates. The gate provides an asymmetric potential of across the transverse direction (y-direction) of the AGNR. Electrons are transported in the x-direction. The plots (a)-(d) show the electron flux at energy, E=0.3eV when E-field is applied in a (a) parallel (P) configuration, i.e. $V_g$=+1V in a homogenous structure. (b) anti-parallel (AP) configuration, i.e. $V_g$=-1V in an homogenous structure. (c) P configuration in a hollowed structure. (d) AP configuration in a hollowed structure. High transmission, T is obtained in (a), (b), and (c). Low T is obtained in (d). The relevant structure parameters are $N_y$=16, $N_x$=100, $N_y'$=2, and $N_x'$=90.

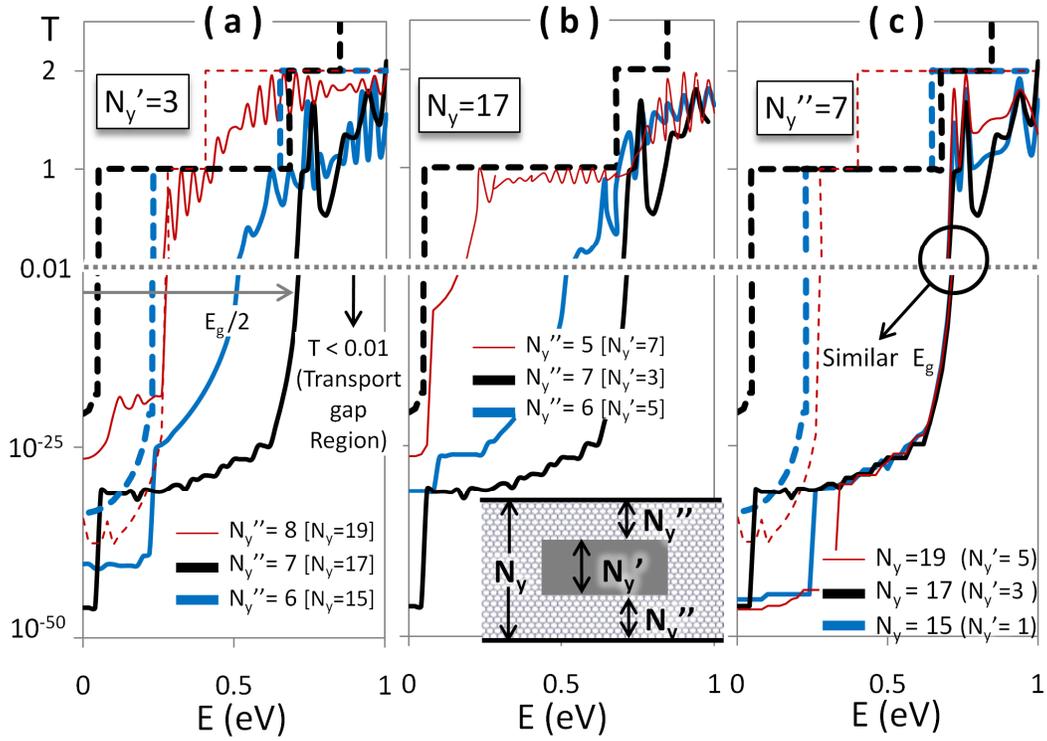

Fig. 4 Transmission-energy T(E) spectrum (a) for different $N_y''$ when $N_y'=3$, (b) for different $N_y''$, when $N_y=17$, and (c) for different $N_y$, when $N_y''=17$, respectively. The dashed lines shows the T(E) of the homogenous AGNR, i.e. $N_y'=0$. $N_y$, $N_y'$, and $N_y''=(N_y-N_y')/2$ are defined in the inset of (b). Note that the plot for T>0.01 (T<0.01) is shown in linear (log) scale.

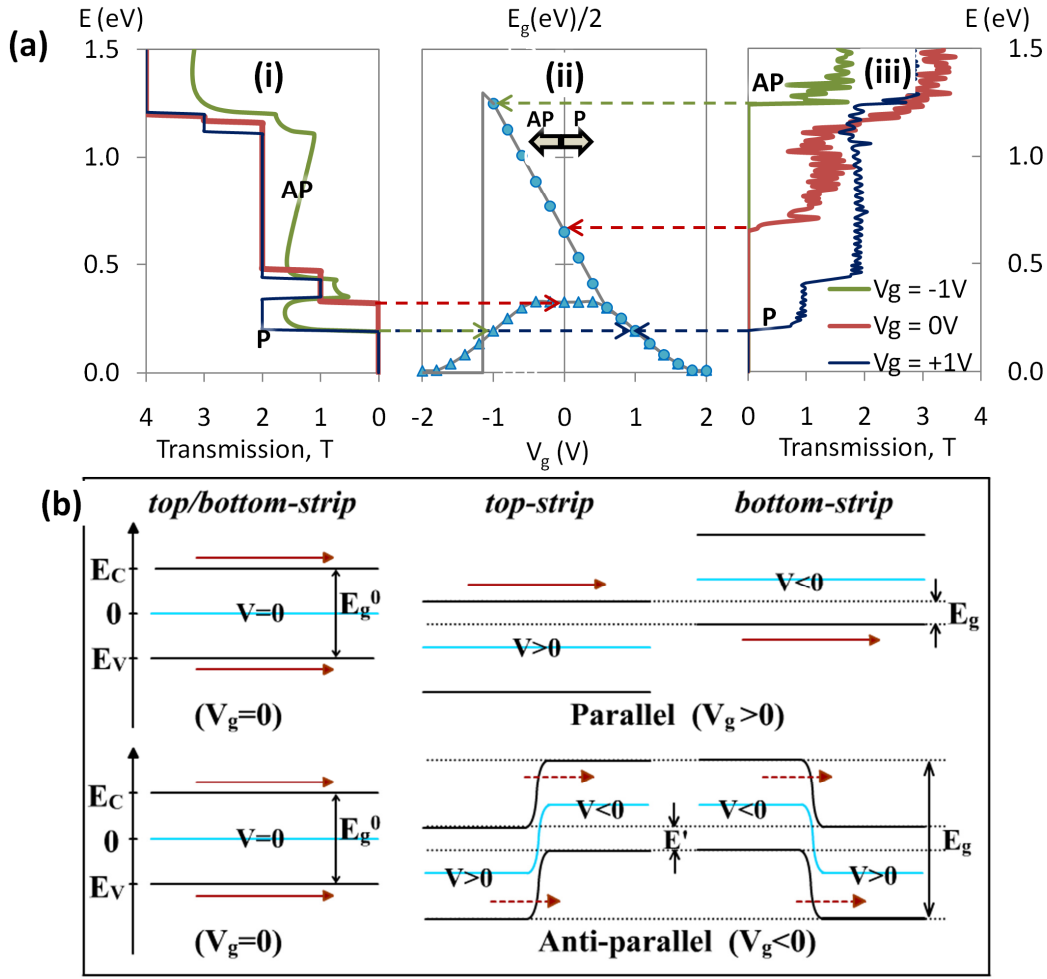

Fig. 5 (a) The central plot shows that the variation of $E_g$ with $V_g$ for homogenous (triangle) and hollowed (circle) GNR. The T(E) curve for under AP (Vg=-1), P (Vg=+1) and Vg=0 for the homogenous and hollowed GNR is shown in the left and the right plot, respectively. [Ny=16, Ny'2, Nx=100, Nx'=90]. (b) Schematics of the effect of Vg on the band structure of the edge strips in the hollowed GNR. When $V_g<0$ ($V_g>0$) the effective $E_g$ is decreased (increased). Note: When the $V_g$ is very low, E' would become negative, resulting in band-to-band tunnelling of electron.

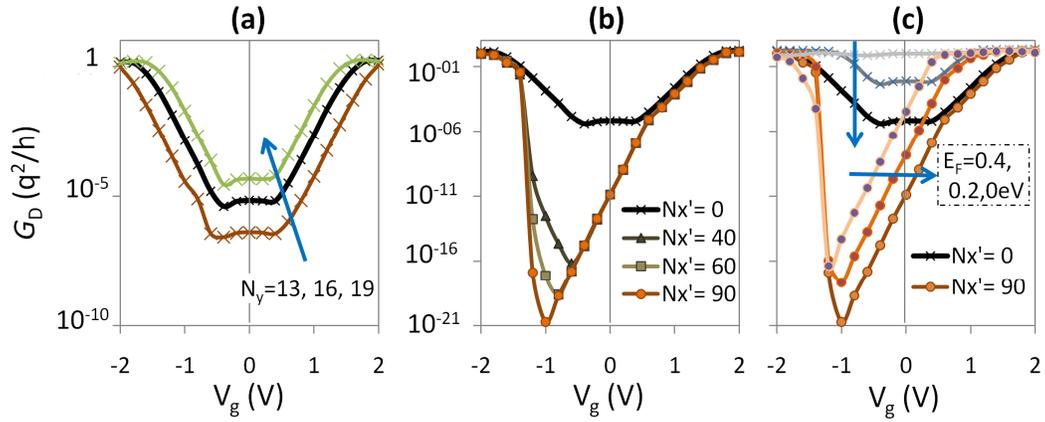

Fig. 6: (a) The source-drain conductance, $G_D$ of homogenous GNRs, as functions of $V_g$ for different GNR width, $N_y$. (b) The $G_D$ of hollowed GNRs, as functions of $V_g$ for different hollow length, $N_x$. (c) The $G_D$ of homogenous (cross) and hollowed GNR (circle), as functions of $V_g$ for varying Fermi level, $E_F$. Unless otherwise stated: $N_y=16$, $N_x'=0$, $E_F=0$eV. All simulations are done at room temperature.